\documentclass[showkeys,showpacs]{revtex4}
\usepackage{graphicx}
\usepackage{amsmath}
\usepackage{amssymb}
\usepackage{color}
\usepackage{enumerate}

\begin{document}
\title{Classical gauge field as a dark matter}

\author{Vladimir Dzhunushaliev}
\email{vdzhunus@krsu.edu.kg}
\affiliation{Department of Physics and Microelectronic
Engineering, Kyrgyz-Russian Slavic University, Bishkek, Kievskaya Str.
44, 720021, Kyrgyz Republic \\ 
Institute of Physicotechnical Problems and Material Science of the NAS
of the Kyrgyz Republic, 265 a, Chui Street, Bishkek, 720071,  Kyrgyz Republic \\
}

\author{Kairat Myrzakulov}
\affiliation{Dept. Gen. Theor. Phys., Eurasian National University, Astana, 010008, Kazakhstan}

\author{Ratbay Myrzakulov}
\affiliation{Dept. Gen. Theor. Phys., Eurasian National University, Astana, 010008, Kazakhstan
}
\email{cnlpmyra1954@yahoo.com}

\begin{abstract}
The model of dark matter is presented where the dark matter is a \emph{classical} gauge field. A spherical symmetric solution of Yang~--~Mills equation is obtained. The asymptotic behavior of the gauge fields and matter density is investigated. It is shown that the distribution of the matter density allows us interpret it as the dark matter. The fitting of a typical rotational curve with the rotational curve created by the spherical solution of SU(3) Yang~--~Mills equation is made. 
\end{abstract}

\date{\today}

\maketitle

\section{Introduction}

One can give such definition for a dark matter \cite{wiki}: ``\ldots dark matter is matter of unknown composition that does not emit or reflect enough electromagnetic radiation to be observed directly, but whose presence can be inferred from gravitational effects on visible matter.'' The nature of the dark matter of the Universe is one of the most challenging problems facing modern physics. Following to L. Smolin \cite{smolin} there exist five great problems in the modern theoretical physics: 
\begin{enumerate}[Problem 1.]
	\item Combine general relativity and quantum theory into a single theory that can claim to be the complete theory of nature. This is called the problem of quantum gravity.
	\item Resolve the problems in the foundations of quantum mechanics, either by making sense of the theory as it stands or by inventing a new theory that does make sense. 
	\item Determine whether or not the various particles and forces can be unified in a theory that explains them all as manifestations of a single, fundamental entity.
	\item Explain how the values of the free constants in the standard model of particle physics are chosen in nature.
	\item Explain dark matter and dark energy. Or, if they don't exist, determine how and why gravity is modified on large scales. More generally, explain why the constants of the standard model of cosmology, including the dark energy, have the values they do.
\end{enumerate}
The problem of the dark matter is the fifth one. 

\begin{figure}[h]
	\centering
		\includegraphics[height=9cm,width=13cm]{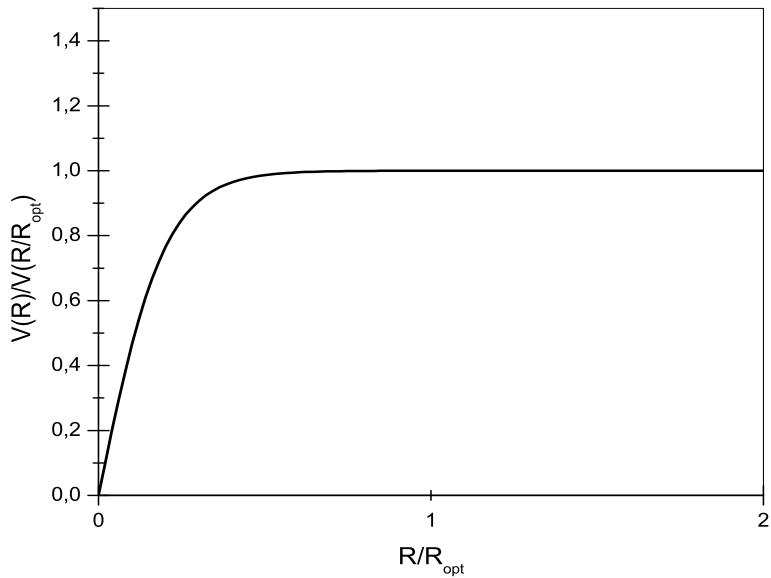}
	\label{rot_curve}
	\caption{Schematical rotation curve of spiral galaxies.}
\end{figure}

Direct observational evidence for dark matter is found from a variety of sources: 
\begin{itemize}
	\item On the scale of galactic halos, the observed flatness of the rotation curves of spiral galaxies is a clear indicator for dark matter.
	\item The measured orbital velocities of galaxies within galactic clusters have been found to be consistent with dark matter observations.
	\item In clusters of galaxies there is a hot intracluster gas. Its temperature allows to measure gravitational potential of a cluster. These data are in agreement with measurements of galaxies speeds and show presence of dark matter. 
	\item The direct evidence of dark matter has been obtained through the study of gravitational lenses. 
\end{itemize}
One of the strongest pieces of evidence for the existence of dark matter is following. Let us consider a rotational velocity $v(r)$ of stars in a galaxy. According to Newton law 
\begin{equation}
	v^2(r) \propto G_N \frac{M(r)}{r} 	
\end{equation}
where $M(r)$ is the mass at a given distance $r$ from the center of a galaxy; $G_N$ is the Newton gravitational constant. The rotational velocity, is measured \cite{fg,rft} by observing 21 cm emission lines in HI regions (neutral hydrogen) beyond the point where most of the light in the galaxy ceases. Schematically a typical rotation curves of spiral galaxies is shown in Fig. \ref{rot_curve} (for details, see Ref. \cite{pss}). 
If the bulk of the mass is associated with light, then beyond the point where most of the light 
stops, $M(r)$ would be a constant and $v(r)^2  \propto 1/r$.  This is not the case, as the rotation
curves appear to be flat, i.e., $v(r) \sim \text{constant}$ outside the core of the galaxy. This implies that $M(r) \propto r$ beyond the point where the light stops. \textcolor{blue}{This fact is the evidence of the existence of dark matter.} The detailed review for the experimental evidence of dark matter can be found in \cite{Olive:2003iq}.

Several categories of dark matter have been postulated: (a) baryonic dark matter; (b) non-baryonic dark matter. The non-baryonic dark matter can be divided into three different types: (b1) hot dark matter - nonbaryonic particles that move ultrarelativistically[15]; (b2) warm dark matter - nonbaryonic particles that move relativistically; (b3) cold dark matter - nonbaryonic particles that move non-relativistically[16]. 

Another approach for the resolution of the dark matter problem is based on a modification of Newton's law or of general relativity, have been proposed to explain the behavior of the galactic rotation curves: (a) a modified gravitational potential \cite{Sa84}; (b) the Poisson equation for the gravitational potential is replaced by another equation \cite{Mi}; (c) alternative theoretical models to explain the galactic rotation curves have been elaborated by Mannheim \cite{Ma93}, Moffat and Sokolov \cite{Mo96} and Roberts \cite{Ro04}; (d) The idea that dark matter is a result of the bulk effects in brane world cosmological models was considered in \cite{Ha}.

All above mentioned approaches to the resolution dark matter problem are based on the assumption that the dark matter is one or another kind of quantum particles. The problem for this approach is that the most of these particles are hypothetical ones: till now these particles were not observed in the nature and in spite of general enthusiasm we do not have any confidence that these particles will be discovered. 

Here we propose the idea that the dark matter is a classical gauge field. This approach is based on the fact that in the consequence of the nonlinearity of Yang~--~Mills equations there exist a spherically symmetric solution without sources (color charge). The matter density in such solution is $\propto r^{-2 \alpha}$ with $\alpha < 2$ that radically differs from the distribution matter density for Coulomb solution. Thus the proposed idea is that some galaxies are immersed in a cloud of a classical gauge field. The SU(3) classical gauge field does not interact with ordinary matter because ordinary matter is colorless. Thus one can suppose that SU(3) gauge field can be invisible matter in galaxies. The problem for such consideration is why the gauge field do not fill all Universe ? The probable answer is that, in certain circumstances, the gauge field goes from a classical phase into a quantum phase. Probably such transition takes place at some distance from the center of the galaxy. 

Let us note that in Ref's \cite{Elizalde:2003ku}-\cite{couple} the similar approach for a dark energy is considered. In Ref. \cite{Elizalde:2003ku} it is shown that the Born-Infeld quantum condensate can play the role of dark energy in the present--day universe. In Ref's \cite{Bamba:2008ja} \cite{Bamba:2008xa} it is demonstrated that both inflation and the late-time acceleration of the universe can be realized in modified Maxwell-$F(R)$ and Yang-Mills-$F(R)$  gravities. In Ref's \cite{z}-\cite{couple} it is supposed that the dark energy is a condensate of Yang~--~Mills gauge field where the effective Lagrangian density of the YM field is calculated up to 1-loop order \cite{pagels} \cite{adler}. With the logarithmic dependence on the field strength, the effective Lagrangian has a form similar to the Coleman-Weinberg scalar effective potential \cite{coleman}. 

\section{Classical gauge theory}

In this section we would like to give a short introduction to SU(N) Yang~--~Mills gauge theory. The corresponding Lagrangian is 
\begin{equation}
	\mathcal L = F^A_{\mu \nu} F^{A \mu \nu}
\label{1-10}
\end{equation}
where $A, B,C = 1,2, \cdots , N$ are the SU(N) color indexes; 
$F^B_{\mu \nu} = \partial_\mu \mathcal A^B_\nu - \partial_\nu \mathcal A^B_\mu + 
g f^{BCD} \mathcal A^C_\mu \mathcal A^D_\nu$ is the field strength; $A^B_\mu$ is the SU(N) gauge potential; $g$ is the coupling constant; $f^{BCD}$ are the structure constants for the SU(N) gauge group. The corresponding Yang~--~Mills field equations are 
\begin{equation}
    \partial_\nu F^{A\mu\nu} = 0.
\label{1-20}
\end{equation}
Here we will consider $N=3$ case of the Yang~--~Mills equations. Equations \eqref{1-20} are nonlinear generalizations of Maxwell equations. The spherally symmetric static solution in electrodynamics is Coulomb potential. Well known spherically symmetric static solution for the SU(2) Yang~--~Mills equations are famous monopole and instanton solutions. The monopole solution has finite energy and is a special solution for corresponding equations. For our goals we will consider practically the same equations as for monopole but with different boundary conditions. Strictly speaking the solutions of equations \eqref{a1-10}--\eqref{a1-60} for \textcolor{blue}{\emph{almost all}} boundary conditions are singular (the energy is infinite) and only for some \textcolor{blue}{\emph{special}} choice of boundary conditions we have regular solution (monopole solution with finite energy). In this work we use solution with infinite energy but we assume that at some distance from the origin the classical phase undergoes a transition to a quantum phase. 

\subsection{Ansatz for the gauge potential}

The classical SU(3) Yang-Mills gauge field $A^B_\mu$ we choose in the following form  \cite{corrigan} 
\begin{eqnarray}
	A_0^2 &=& - 2 \frac{z}{gr^2} \chi(r), 
\label{a1-10}\\
	A_0^5 &=& 2 \frac{y}{gr^2} \chi(r), 
\label{a1-12}\\
	A_0^7 &=& - 2 \frac{x}{gr^2} \chi(r), 
\label{a1-14}\\
	A^2_i &=& 2 \frac{\epsilon_{3ij} x^j}{gr^2} \left[ h(r) + 1 \right] ,
\label{a1-20}\\
	A^5_i &=& -2 \frac{\epsilon_{2ij} x^j}{gr^2} \left[ h(r) + 1 \right] ,
\label{a1-30}\\
	A^7_i &=& 2 \frac{\epsilon_{1ij} x^j}{gr^2} \left[ h(r) + 1 \right] 
\label{a1-40}
\end{eqnarray}
here $A^{2,5,7}_\mu \in SU(2) \subset SU(3)$; the SU(2) is the subgroup of SU(3); $\mu =0,1,2,3$ is the spacetime index and $i,j,k = 1,2,3$ are space indexes. The remaining components are belong to the coset $SU(3)/SU(2)$
\begin{eqnarray}
	\left( A_0 \right)_{\alpha , \beta} &=& 2 \left( 
		\frac{x^\alpha x^\beta}{r^2} - \frac{1}{3} \delta^{\alpha \beta}
	\right) \frac{w(r)}{gr} ,
\label{a1-50}\\
	\left( A_i \right)_{\alpha \beta} &=& 2 \left(
		\epsilon_{is \alpha} x^\beta + \epsilon_{is \beta} x^\alpha
	\right) \frac{x^s}{gr^3} v(r) ,
\label{a1-60}
\end{eqnarray}
here $\epsilon_{ijk}$ is the absolutely antisymmetric Levi-Civita tensor. The coset components $\left( A_\mu \right)_{\alpha \beta}$ are written in the matrix form and by definition are  
\begin{equation}
	\left( A_\mu \right)_{\alpha \beta} = 
	\sum \limits_{m=1,3,4,6,8} A_\mu^m \left( T^m \right)_{\alpha , \beta} 
\label{a1-70}
\end{equation}
where $T^B = \frac{\lambda^B}{2}$ are the SU(3) generators, $\lambda^B$ are the Gell-Mann matrices. 

\subsection{Yang~-~Mills equations}

The corresponding Yang~-~Mills equations \eqref{1-20} with the potential \eqref{a1-10}~--~\eqref{a1-60} and ($\chi(r) = h(r) = 0$) are 
\begin{eqnarray}
	x^2 w'' &=& 6w v^2 ,
\label{a2-10}\\
	x^2 v'' &=& v^3 - v - v w^2 
\label{a2-20}
\end{eqnarray}
here we introduce the dimensionless radius $x = r/r_0$, $r_0$ is an arbitrary constant. The asymptotic behavior of the functions $v(x), w(x)$ by $x \gg 1$ are \cite{Dzhunushaliev:1999fy} 
\begin{eqnarray}
	v(x) &\approx& A \sin \left( x^\alpha + \phi_0 \right),
\label{a2-25}\\
	w(x) &\approx& \pm \left[
		\alpha x^\alpha + \frac{\alpha - 1}{4} 
		\frac{\cos \left( 2 x^\alpha  + 2 \phi_0 \right)}{x^\alpha}
	\right] ,
\label{a2-30}\\
	3 A^2 &=& \alpha (\alpha - 1)
\label{2-40}
\end{eqnarray}
with $\alpha > 1$. The energy density $\epsilon(x)$ is 
\begin{equation}
	\epsilon(r) = 
	- F^a_{0i} F^{a0i} + \frac{1}{4} F^a_{ij} F^{aij} = 
	\frac{1}{g^2 r_0^4} \left[ 
	4 \frac{{v'}^2}{x^2} + 
	\frac{2}{3} \frac{\left( x w' - w \right)^2}{x^4} + 
	2 \frac{\left( v^2 - 1 \right)^2}{x^4} + 
	4 \frac{ v^2 w^2}{x^4} 
	\right] = 
	\frac{1}{g^2 r_0^4} \varepsilon(x).
\label{a2-50}
\end{equation}
Using the asymptotic behavior of the gauge potential \eqref{a2-25} \eqref{a2-30} the asymptotic behavior of the energy density is 
\begin{equation}
	\varepsilon_\infty(x) \approx 
	\frac{2}{3} 
	\alpha^2 \left( \alpha - 1 \right) \left( 3 \alpha - 1 \right) 
	\left( \frac{r}{r_0} \right)^{2 \alpha - 4}. 
\label{a2-60}
\end{equation}

\subsection{Numerical investigation}
\label{numerical}

\begin{figure}[h]
\begin{minipage}[t]{.45\linewidth}
  \begin{center}
  \fbox{
  \includegraphics[height=5cm,width=7cm]{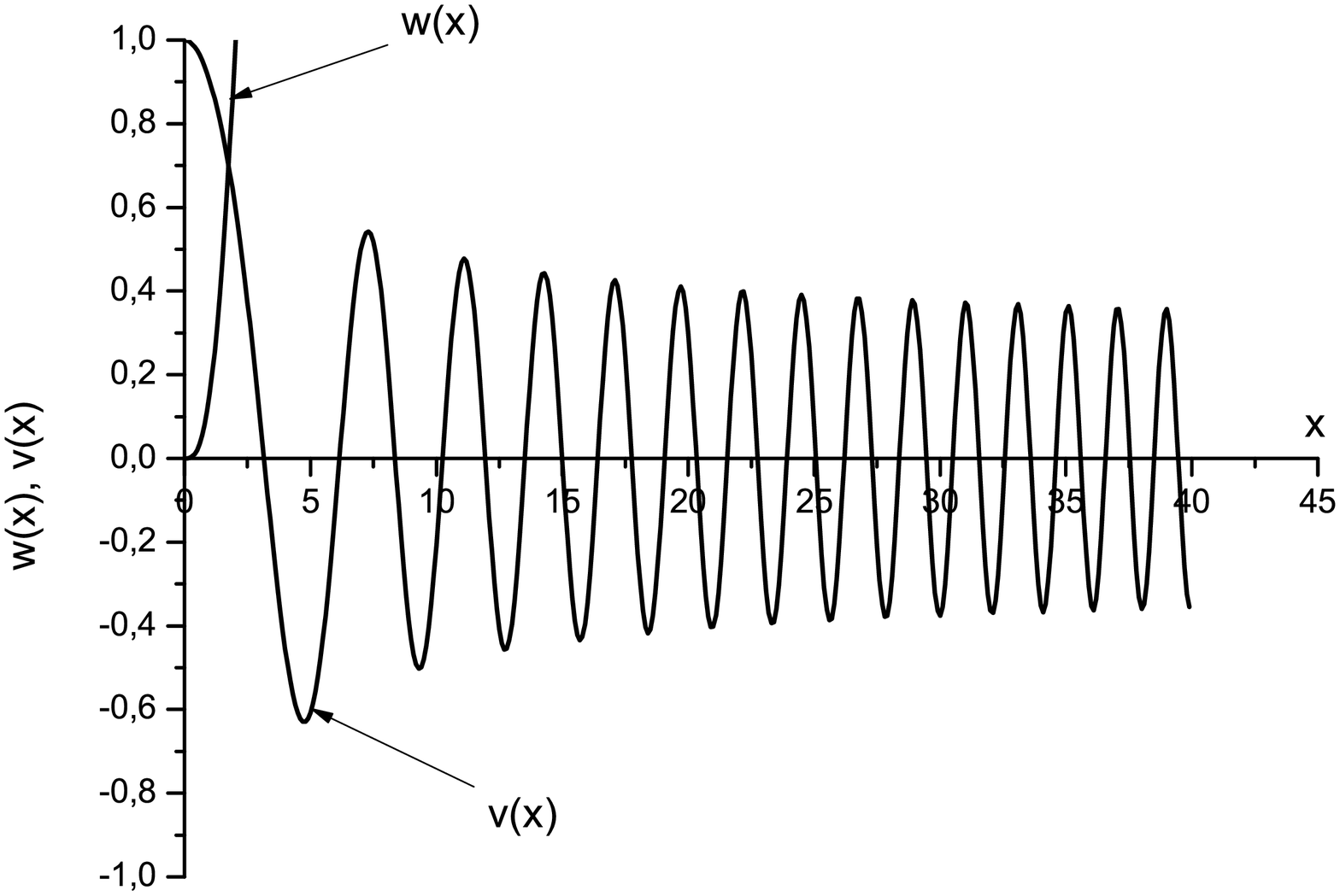}}
  \caption{The profiles of functions $w(x), v(x))$, $v_2=-0.2$, $w_3=1$.}
  \label{fg1}
  \end{center}
\end{minipage}\hfill
\begin{minipage}[t]{.45\linewidth}
  \begin{center}
  \fbox{
  \includegraphics[height=5cm,width=7cm]{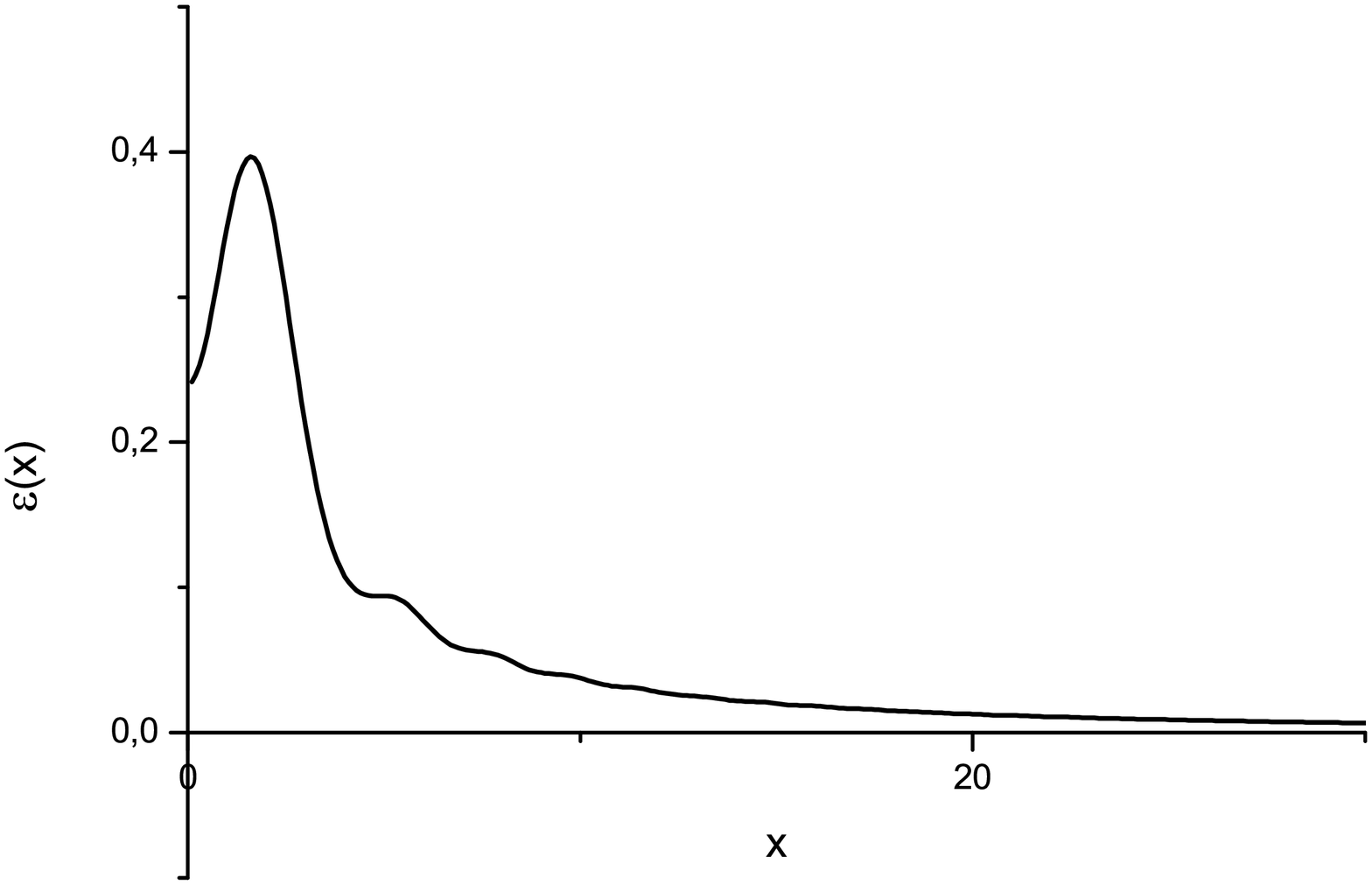}}
  \caption{The profile of the	dimensionless energy density  
  $\varepsilon (x)$.}
  \label{fg2}
  \end{center}
\end{minipage}\hfill 
\end{figure}

In this subsection we present the typical numerical solution of Eq's \eqref{a2-10} \eqref{a2-20}. In the consequence of occurrence of the factor $x^2$ in the front of left hide side equations \eqref{a2-10} \eqref{a2-20} we have to start the numerical calculations not from $x=0$ but from from $x = \delta \ll 1$. To do that we should have an approximate solution near to the origin 
\begin{eqnarray}
	v(r) &=& 1 + \frac{1}{2} \left( r_0^2 v_2 \right) \left( \frac{r}{r_0} \right)^2 + 
	\mathcal O \left[ \left( \frac{r}{r_0} \right)^4 \right] = 
	1 + v'_2 \frac{x^2}{2}  + \mathcal O \left( x^4 \right) , 
\label{a3-10}\\
	w(r) &=&  \frac{1}{6} \left( r_0^3 w_3 \right) \left( \frac{r}{r_0} \right)^3 + 
	\mathcal O \left[ \left( \frac{r}{r_0} \right)^5 \right] = 
	w'_3 \frac{x^3}{6} + \mathcal O \left( x^5 \right).
\label{a3-20}
\end{eqnarray}
Therefore the natural choice of the parameter $r_0$ is 
\begin{equation}
	\text{either} \quad r_0^2 = \frac{1}{v_2} \quad
	\text{or} \quad 
	r_0^3 = \frac{1}{w_3}.
\label{a3-30}
\end{equation}
For the numerical calculations we choose the parameter $r_0$ as 
\begin{equation}
	r_0 = \frac{1}{w_3^{1/3}}.
\label{a3-40}
\end{equation}
The typical behavior of functions $v(x)$ and $w(x)$ is presented in Fig.~\ref{fg1} that is in agreement with the asymptotic behavior \eqref{a2-25} \eqref{a2-30}. 

The mass density $\rho(r)$ is 
\begin{equation}
	\rho(r) = \frac{1}{g^2 c^2 r_0^4} \rho(x)
\label{a3-50}
\end{equation}
where $\rho(x) = \varepsilon(x)$ and $\varepsilon(x)$ is given in Eq.~\eqref{a2-50}. The profile of the dimensionless energy density  $\varepsilon(x)$ in Fig.~\ref{fg2} is presented.

\subsection{The comparison with a Universal Rotation Curve of spiral galaxies}

The idea presented in this work is that in a galaxy there exists an ordinary visible matter (barionic matter which glows) and an invisible matter (classic gauge field which does not interact with electromagnetic waves). The visible matter is immersed into a cloud of the classical gauge field. The main goal of this work is to compare the experimental rotational curve for stars in the galaxy outside light core with the rotational curve created by the distribution of the classical gauge field. 

In Ref.~\cite{Persic:1995ru} a Universal Rotation Curve of spiral galaxies is offered that describes any rotation curve at any radius with a very small cosmic variance
\begin{eqnarray}
	\left[ \frac{V^2_{URC} \left( \frac{r}{R_{opt}} \right)}{V(R_{opt})} \right]^2 &=& 
	 \left( 0.72 + 0.44 \log \frac{L}{L_*} \right) 
		\frac{1.97 X^{1.22}}{ \left( X^2 + 0.78^2 \right)^{1.43}} + 
		1.6\, e^{-0.4(L/L_*)} \frac{X^2}{X^2 + 1.5^2 
		\left( \frac{L}{L_*} \right)^{0.4}} = 
\nonumber\\
	&& V_{LM}^2 + V_{DM}^2	, \quad \text{Km$^2$/s$^2$}
\label{5b-10}\\
	\left[ \frac{V^2_{LM} \left( \frac{r}{R_{opt}} \right)}{V(R_{opt})} \right]^2 &=& 
	\left( 0.72 + 0.44 \log \frac{L}{L_*} \right) 
		\frac{1.97 X^{1.22}}{ \left( X^2 + 0.78^2 \right)^{1.43}},
\label{5b-14}\\
	\left[ \frac{V^2_{DM} \left( \frac{r}{R_{opt}} \right)}{V(R_{opt})} \right]^2 &=& 
	1.6\, e^{-0.4(L/L_*)} \frac{X^2}{X^2 + 1.5^2 
		\left( \frac{L}{L_*} \right)^{0.4}}
\label{5b-18}
\end{eqnarray}
where $R_{opt} \equiv 3.2\,R_D$ is the optical radius and $R_D$ is the disc exponential length-scale; $X = r/R_{opt}$; $L$ is the luminosity; the first term $V_{LM}^2$ is the rotation curve for the light matter and the second term $V_{DM}^2$ the rotation curve for the dark matter. Our goal is to compare the rotation curve for the color fields 
\begin{equation}
	V^2(r) = G_N \frac{\mathcal{M}(r)}{r} = 
	4 \pi G_N \frac{1}{r}
	\int \limits^r_0 r^2 \rho(r) dr = 
	\frac{G_N \hbar}{c^3} \frac{c^2}{{g'}^2 r_0^2} \frac{\mathcal{M}(x)}{x} = 
	\left[
		\frac{1}{{g'}^2} \left( \frac{l_{Pl}}{r_0} \right)^2 
		\frac{\mathcal{M}(x)}{x}
	\right] c^2 
\label{5b-20}
\end{equation}
where $\mathcal{M}(x)$ is the dimensionless mass of the color fields $A^B_\mu$ inside the sphere of radius $r = x r_0$; ${g'}^2 = g^2 c \hbar/ 4\pi$ is the dimensionless coupling constant with the experimental rotation curve for the dark matter \eqref{5b-18} where, for example, $L/L_* = 1$
\begin{equation}
	V_{DM} \left( \frac{r}{R_{opt}} \right) = 
	V(R_{opt}) \left( 
		\frac{1.07251 \; X^2}{X^2 + 1.5^2} 
	\right)^{1/2} {\rm Km/s}.
\label{5b-30}
\end{equation}
\begin{figure}[h]
\begin{minipage}[t]{.45\linewidth}
  \begin{center}
  \fbox{
  \includegraphics[height=5cm,width=7cm]{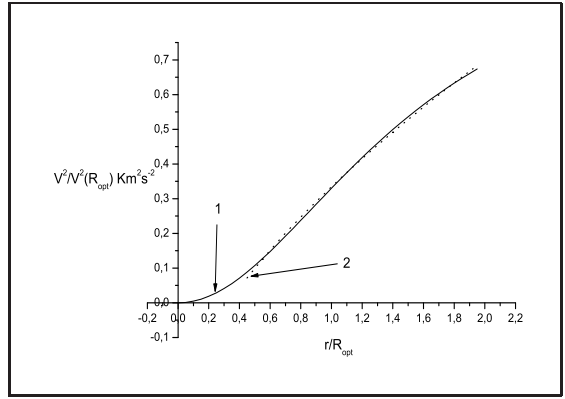}}
	  \caption{The comparison of DM rotation curve \eqref{5b-30} (curve~1) with the rotation curve \eqref{3-78} (curve~2) for the SU(3) classical color field \eqref{a1-10}-\eqref{a1-40}. 
	  $\alpha \approx 1.31, g' = 1, R_{opt} = 20 KPs = 6*10^{17} m, V_{opt} = 100 Km/s$.}
  \label{fg4}
  \end{center}
\end{minipage}\hfill 
\end{figure}
\par
Far away from the center the dimensionless energy density $\varepsilon_\infty(x)$ is presented in equation~\eqref{a2-60}. In order to fit equation~\eqref{a2-60} we break up equation \eqref{5b-20} into two terms 
\begin{eqnarray}
	V^2 &=& c^2 \frac{1}{{g'}^2} \left( \frac{l_{Pl}}{r_0^2}^2 \right) \frac{1}{x} 
	\left( 
		\int \limits_0^{x} x^2 \left[ \varepsilon(x) - \varepsilon_\infty(x) \right] dx + 
		\int \limits_{x_1}^x x^2 \varepsilon_\infty(x) dx
	\right) 
\nonumber \\
	&\approx& \left[
		c^2 \frac{1}{{g'}^2} \left( \frac{l_{Pl}}{r_0^2}^2 \right) \frac{1}{x} 
		\int \limits_0^x x^2 \varepsilon_\infty(x) dx 
	\right] - V^2_0 
\label{3-73}
\end{eqnarray}
where 
\begin{equation}
	V^2_0 = c^2 \frac{1}{{g'}^2} \left( 
	\frac{l_{Pl}}{r_0^2}^2 \right) \frac{1}{x}  \int \limits_0^{x} x^2 
	\left[ \varepsilon_\infty(x) - \varepsilon(x) \right] dx .
\label{3-76}
\end{equation}
The numerical value of $V^2_0$ is defined near to the center of galaxy where according to Eq. \eqref{3-76} the difference $\varepsilon_\infty(x) - \varepsilon(x)$ is maximal. Thus the asymptotic behavior of the rotation curve for the domain filled with the SU(3) gauge field is 
\begin{equation}
	V^2 \approx \left[ 
	\frac{2}{3} \frac{1}{{g'}^2} \left( \frac{l_{Pl}}{r_0} \right)^2 
		\frac{\alpha^2 \left( \alpha - 1 \right) \left( 3 \alpha - 1 \right) }{2 \alpha - 1}
		\left( \frac{r}{r_0} \right)^{2 \alpha - 2} 
	\right]c^2 - V^2_0 .
\label{3-78}
\end{equation}
In Fig.~\ref{fg4} the profiles of the dark matter rotational curve \eqref{5b-18} and fitting curve \eqref{3-78} are presented. The value of parameter $\alpha$ is given from section~\ref{numerical}. The details of fitting 
\begin{eqnarray}
	r_0 &\approx & 2.01 \cdot 10^{-18} \; \text{cm}, 
\label{3-81}\\
	V_0 &\approx & 32.25 
	\text{ Km}/\text{s}
\label{3-91}
\end{eqnarray}
in Appendix~\ref{asymptotic} are presented. 

\section{Cut-off the region filled with classical gauge field}

The gauge field distribution considered here has an infinite energy in the consequence of the asymptotic behavior \eqref{a2-60} of the energy density \eqref{a2-50}. Consequently it is necessary to have a mechanism to cut-off the distribution of the classical gauge fields in the space. In our opinion it can happen in the following way. At some distance from the origin the gauge field undergoes the transition from the classical state to quantum one and the quantum field tends very fast to its vacuum expectation value. It is very important to note that the gauge field in the vacuum state must be described in a nonperturbative manner. 

The physical reason why such transition takes place is following. As we see from Eq.\eqref{a2-25} \eqref{a2-30} and Fig.\ref{fg1} asymptoticly the gauge potentials are oscillating functions with increasing frequency. Far away from the origin the frequency is so big that it is necessary quantum fluctuations take into account. In this way the transition from the classical state to quantum one takes place. 

To estimate a transition radius we follow to the Heisenberg uncertainty principle 
\begin{equation}
	\frac{1}{c} \; \Delta F^a_{ti} \; \Delta A^{a i} \; \Delta V \approx \hbar
\label{3-10}
\end{equation}
here $\Delta F^a_{ti}$ is a quantum fluctuation of the color electric field $F^a_{ti}$; $\Delta A^{a i}$ is a quantum fluctuation of the color electric potential $A^{a i}$; $\Delta V$ is the volume where the quantum fluctuations $\Delta F^a_{ti}$ and $\Delta A^{a i}$ takes place and not any summation over repeating indexes. 
\par 
For the ansatz \eqref{a1-50}-\eqref{a1-60}
\begin{equation}
	F^2_{t \theta} = - \frac{2 \sin \theta}{g} \frac{v w}{r} .
\label{3-20}
\end{equation}
Let us to introduce the physical component of the $F^2_{t \theta}$ 
\begin{equation}
	\left| \tilde F^2_{t \theta} \right| = 
	\sqrt{F^2_{t \theta} F^{2\theta}_t} = \frac{2 \sin \theta}{g} \frac{v w}{r^2} .
\label{3-30}
\end{equation}
To an accuracy of a numerical factor the fluctuations of the SU(3) color electric field are 
\begin{equation}
	\Delta \tilde F^2_{t \theta} \approx \frac{1}{g} \frac{1}{r^2} \left(
		\Delta v \; w + v \; \Delta w
	\right).
\label{3-40}
\end{equation}
For the ansatz \eqref{a1-50}-\eqref{a1-60}
\begin{eqnarray}
	A^2_\theta = 0,
\label{3-70}\\
	A^{1,3,4,6,8}_\theta \approx \frac{1}{g} v.
\label{3-80}
\end{eqnarray}
The physical components of the gauge potential $A^{1,3,4,6,8}_\theta$ 
\begin{equation}
	\left| \tilde A^{1,3,4,6,8}_\theta \right| = 
	\sqrt{A^{1,3,4,6,8}_\theta A^{1,3,4,6,8; \theta}} \approx 
	\frac{1}{g} \frac{v}{r} .
\label{3-90}
\end{equation}
Now we assume that the quantum fluctuations $\Delta \tilde A^2_\theta$ of the component 
$\tilde A^2_\theta$ have the same order as the quantum fluctuations of the components 
$\tilde A^{1}_\theta$
\begin{equation}
	\Delta \tilde A^2_\theta \approx \Delta \tilde A^1_\theta \approx 
	\frac{1}{g} \frac{\Delta v}{r} .
\label{3-100}
\end{equation}
The volume $\Delta V$ where supposed quantum fluctuations take place is 
\begin{equation}
	\Delta V = 4\pi r^2 \Delta r .
\label{3-110}
\end{equation}
The period of space oscillations by $r \gg r_0$ can be defined in the following way 
\begin{equation}
	\left( x + \lambda \right)^\alpha - x^\alpha \approx 
	\alpha \frac{\lambda}{x^{1 - \alpha}} = 2\pi; 
	\quad x = \frac{r}{r_0}.
\label{3-120}
\end{equation}
We suppose that the place where the SU(3) classical color field becomes quantum one is defined as the place where the quantum fluctuations in the volume $\Delta V = 4\pi r^2 \Delta r$ with 
\begin{equation}
	\frac{\Delta r}{r_0} \approx \lambda 
	\approx \frac{1}{\alpha} \frac{2 \pi}{x^{\alpha - 1}}
\label{3-130}
\end{equation}
of the corresponding field becomes comparable with magnitude of these fields 
\begin{equation}
	\Delta v \approx v, \quad 
	\Delta w \approx w
\label{3-140}
\end{equation}
Substituting of Eq's \eqref{3-40}, \eqref{a2-25}-\eqref{a2-30}, \eqref{3-100}, \eqref{3-110} , \eqref{3-130} and \eqref{3-140} into Eq.~\eqref{3-10} we obtain 
\begin{equation}
	\left( \frac{g'}{A} \right)^2 \approx 2 \pi
\label{3-150}
\end{equation}
where $\frac{1}{{g'}^2} = \frac{4 \pi}{g^2 \hbar c}$ is the dimensionless coupling constant that is similar to the fine structure constant in quantum electrodynamics $\alpha = \frac{e^2}{\hbar c}$. In quantum chromodynamics $\beta = 1/{g'}^2 \geq 1$. If we choose $1/{g'} \approx 1$ and from Fig.~\ref{fg1} we take $A \approx 0.4$ we see that 
\begin{equation}
	\left( \frac{g'}{A} \right)^2 \approx 6.25 
\label{3-160}
\end{equation}
that is comparable with $2 \pi \approx 6.28$. 
\par 
Thus in this section we have shown that if the condition \eqref{3-150} is true then at some distance from the center the transition from the classical phase to quantum one occurs. Unfortunately the rough estimation presented in this section does not allow us to calculate the radius where such transition takes place. For the exact evaluation of the place where such transition happens it is necessary to have \emph{non-perturbative} quantization methods which are missing at the moment. 

\section{Invisibility of the color dark matter}

The main question in any dark matter model is its invisibility. For the model presented here the answer is very simple: the SU(3) color matter (dark matter in this context) is invisible because color gauge fields interact with color charged particles only. But at the moment in the nature we do not know any particles with SU(3) color charge. In principle such particles can be SU(3) monopoles but up to now the monopoles are not experimentally registered. 

For more details we write SU(3) Lagrangian interacting with matter 
\begin{equation}
	\mathcal L_{QCD} = - \frac{1}{2} \mathrm{tr} F^A_{\mu \nu} F^{A \mu \nu} + 
	\sum \limits_k^{n_f} \bar q \left( 
		i \gamma^\mu D_\mu - m_k
	\right) q_k, 
\label{4-10}
\end{equation}
where 
\begin{eqnarray}
	D_\mu q_k &=& \left(\partial_\mu - i g A_\mu  \right) q_k ,
\label{4-20}\\
	A_\mu &=& \sum \limits_{B=1}^{8} A^B_\mu \frac{\lambda^B}{2}
\label{4-30}
\end{eqnarray}	
$D_\mu$ is the covariant derivative; $A_mu$ is gauge potential in the matrix form. From the term 
$i g \bar q A_\mu q$ in equation \eqref{4-20} we see that the SU(3) color field has an interaction with quarks only. But the quarks are not observable in the nature. The baryon matter is colorless in the consequence with the confinement of quarks in hadrons and therefore the color dark matter does not interact with the light (photons). As we see above the gauge dark matter can be seen in during of its gravitational field. 

Thus it is interesting that the problem of the dark matter in astrophysics is connected with the problem of confinement in high energy physics. 

\section{Conclusions}

In this work we have suggested the idea that the dark matter model is SU(3) gauge field. We have shown that in SU(3) Yang~--Mills theory there exists a spherical symmetric distribution of the gauge potential with slow decreasing matter density. The asymptotic behavior of the density allow us to describe the rotational curve for the stars in elliptic galaxies. The fitting of the typical rotational curve gives us parameters which have a good agreement with the parameters of above mentioned spherical solution of Yang~--Mills equations.

\section*{Acknowledgements}

V.D. is grateful to the Research Group Linkage Programme of the Alexander von Humboldt Foundation for the support of this research. 
\appendix 

\section {Heisenberg's quantization of strongly interacting fields applying for gauge fields} 

In this section we would like to bring some arguments that nonperturbative quantized SU(3) gauge field tends very quickly to zero. To do so we use the Heisenberg approach \cite{heisenberg} for a nonperturbative quantization of a nonlinear spinor field. In quantizing strongly interacting SU(3) gauge fields - via Heisenberg's non-perturbative method \cite{heisenberg} one first replaces the classical fields by field operators 
$A^B_{\mu} \rightarrow \widehat{A}^B_\mu$. This yields the
following differential equations for the field operators
\begin{equation}
    \partial_\nu \widehat {F}^{B \mu\nu} = 0.
\label{sec2-31}
\end{equation}
These nonlinear equations for the field operators of the nonlinear quantum fields can be used to determine expectation values for the field operators $\widehat {A}^B_\mu$. Starting with Eq. \eqref{sec2-31} one can generate an operator differential equation for the product 
$\widehat {A}^B_\mu \widehat {A}^C_\nu$ consequently allowing the determination of the Green's function $\mathcal{G}^{BC}_{\mu\nu}$
\begin{eqnarray}
	\mathcal{G}^{BC}_{\mu\nu} (x,y) &=& 
	\left\langle Q \left|
	  \widehat {A}^B_\mu(x) \widehat {A}^C_\nu (y)
  \right| Q \right\rangle , 
\label{sec2-32}\\
  \left\langle Q \left|
	  \widehat {A}^B(x) \partial_{y\nu} \widehat {F}^{C\mu\nu}(x)
  \right| Q \right\rangle &=& 0
\label{sec2-33}
\end{eqnarray}
where $\left. \left. \right| Q \right\rangle$ is a quantum state. However this equation will in it's turn contain other, higher order Green's functions. Repeating these steps leads to an infinite set of equations connecting Green's functions of ever increasing order. Let us note that absolutely similar idea is applied in turbulent hydrodynamics for correlation functions all orders. This construction, leading to an infinite set of coupled, differential equations, does not have an exact, analytical solution and so must be handled using some approximation. The basic approach in this case is to give some physically reasonable scheme for cutting off the infinite set of equations for the Green's functions. 

One can use Heisenberg's approach to reduce the initial SU(3) Lagrangian to an effective Lagrangian describing two interacting scalar fields (for details see Ref. \cite{Dzhunushaliev:2006di}). Two scalar fields $\phi$ and $\chi$ appear in such approach. We assume that in the first approximation two points Green's functions are bilinear combinations of scalar fields $\phi$ and $\chi$
\begin{eqnarray}
  \left\langle A^a_i (x) A^b_j (y) \right\rangle &=& c_{mn} \chi^m(x) \chi^n(y),
\label{sec2-34}\\
  \left\langle A^m_i (x) A^n_j (y) \right\rangle &=& d_{ab} \phi^a(x) \phi^b(y),
\label{sec2-35}\\
	\left\langle A^m_0 (x) A^n_0 (y) \right\rangle & \approx & 0
\label{sec2-36}
\end{eqnarray}
where $A^a_\mu \in SU(2) \subset SU(3), a=1,2,3$; $A^m_\mu \in SU(3)/SU(2), m = 4,5,6,7,8$; $c_{mn},d_{ab}$ are some coefficients. 

The next step is the calculation of the 4-points Green's functions. It is assumed that they are a bilinear combination of 2-- points Green functions 
\begin{eqnarray}
  \left\langle A^m_\mu(x) A^n_\nu(y) A^p_\alpha(z) A^q_\beta(u) \right\rangle &=&
  \lambda_1 \left[ \mathcal{G}^{mn}_{\mu\nu} \mathcal{G}^{pq}_{\alpha \beta} + 
	  \frac{\mu_1^2}{4} \left(
	  	\delta^{mn} \eta_{\mu \nu} \mathcal{G}^{pq}_{\alpha \beta} + 
	  	\delta^{pq} \eta_{\alpha \beta} \mathcal{G}^{mn}_{\mu \nu}
	  \right) + 
	  \frac{\mu_1^4}{16} 
	  \delta^{mn} \eta_{\mu \nu} \delta^{pq} \eta_{\alpha \beta} 
	  \right] + 
\nonumber \\
	  &&
	  ( \text{permutations of indices} )
\label{sec2-37}
\end{eqnarray}
the same for remaining indexes 
\begin{eqnarray}
  \left\langle A^a_\mu(x) A^b_\nu(y) A^c_\alpha(z) A^d_\beta(u) \right\rangle &=&
  \lambda_2 \left[ \mathcal{G}^{ab}_{\mu\nu} \mathcal{G}^{cd}_{\alpha \beta} + 
	  \frac{\mu_2^2}{4} \left(
	  	\delta^{ab} \eta_{\mu \nu} \mathcal{G}^{cd}_{\alpha \beta} + 
	  	\delta^{cd} \eta_{\alpha \beta} \mathcal{G}^{ab}_{\mu\nu} \right) + 
	  \frac{\mu_2^4}{16} \delta^{ab} \eta_{\mu \nu} \delta^{cd} \eta_{\alpha \beta}
	  \right] + 
\nonumber \\
	  &&
	 \text{(permutations of indices)}
\label{2f1-30}
\end{eqnarray}
with some constants $\lambda_{1,2}, \mu_{1,2}$. The assumptions \eqref{sec2-33}-\eqref{2f1-30} allows us to average the SU(3) Lagrangian 
\begin{equation}
  \mathcal{L}_{SU(3)} = - \frac{1}{4}	F^A_{\mu \nu} F^{A \mu \nu}, 
\label{2f-40}
\end{equation}
and bring it to the form 
\begin{eqnarray}
  \mathcal{L}_{eff} &=& \left\langle  \mathcal{L}_{SU(3)} \right\rangle = 
  \frac{1}{2} \left( \partial_\mu \phi^a \right) 
    \left( \partial^\mu \phi^a \right) 
    + \frac{1}{2}  \left( \partial_\mu \chi^m \right) 
    \left( \partial^\mu \chi^m \right) - V(\phi^a, \chi^m) ,
\label{2f-50}\\
	V(\phi^a, \chi^m) &=& \frac{\lambda_1}{4} \left(
        \phi^a \phi^a - \mu_1^2 
    \right)^2 - 
    \frac{\lambda_2}{4} \left(
        \chi^m \chi^m - \mu_2^2 
    \right)^2 - \frac{1}{2} \left( \phi^a \phi^a \right) \left( \chi^m \chi^m \right) 
\label{2f-55}
\end{eqnarray}
with the field equations 
\begin{eqnarray}
	\nabla_\mu \left( 
		\nabla^\mu \phi^a
	\right) &=& - \frac{\partial V\left( \phi^a, \chi^m \right)}{\partial \phi^a} ,
\label{sec2f-60}\\
	\nabla_\mu \left( 
		\nabla^\mu \chi^m 
	\right) &=& - \frac{\partial V\left( \phi^a, \chi^m \right)}{\partial \chi^m} .
\label{sec2f-70}
\end{eqnarray}
Let us consider the spherically symmetric case $\phi^a = k \phi(r), \chi^m = k \chi(r)$ where $k$ is some constant. In this case the field equations are 
\begin{eqnarray}
	\frac{d^2 \phi}{d r^2} + \frac{2}{r} \frac{d \phi}{d r} &=& \phi \left[
		\chi^2 + \lambda_1 \left( \phi^2 - \mu_1^2 \right)
	\right] ,
\label{sec2f-80}\\
	\frac{d^2 \chi}{d r^2} + \frac{2}{r} \frac{d \chi}{d r} &=& \chi \left[
		\phi^2 + \lambda_2 \left( \chi^2 - \mu_2^2 \right)
	\right] .
\label{sec2f-90}
\end{eqnarray}
It is easy to see that asymptoticly the solution has the form 
\begin{eqnarray}
	\phi(x) &\approx& m_1 + 
	\phi_\infty \frac{e^{-\left( r - r_q \right) \sqrt{2 \lambda_1 \mu_1^2}}}{r}  ,
\label{sec3-30}\\
	\chi(x) &\approx& \chi_\infty \frac{e^{-\left( r - r_q \right) \sqrt{m_1^2 - \lambda_2 \mu_2^2}}}{r} 
\label{sec3-40}
\end{eqnarray}
where $\phi_\infty, \chi_\infty, r_q$ are constants. We suppose that this solution describes the non-perturbative quantized SU(3) gauge field after the transition from classical phase to quantum one occurs. We see that the non-perturbative quantized gauge field decreases very quickly (exponentially) after transition to the quantum phase and consequently the total mass becomes finite one. 

\section{Fitting of rotational curve of gauge field}
\label{asymptotic}

For the fitting of the rotational curve \eqref{3-78} we use the data from the Universal Rotational Curve \eqref{5b-30}. The fitting equation is equation \eqref{3-78} in the form 
\begin{equation}
	\frac{V^2}{V^2_{opt}} = A x^B + C, 
\label{a3-02}
\end{equation}
where
\begin{eqnarray}
	A &=& \frac{2}{3} 
	\frac{\alpha^2 \left( \alpha - 1 \right) \left( 3 \alpha - 1 \right) }{2 \alpha - 1}
	\frac{c^2}{V^2_{opt}}
	\frac{1}{{g'}^2} \left( \frac{l_{Pl}}{r_0} \right)^2 
	\left( \frac{R_{opt}}{r_0} \right)^{2 \alpha - 2} 
\label{a3-04}\\
	B &=& 2 \alpha - 2, 
\label{a3-06}\\
	C &=& - V_0^2
\label{a3-08}
\end{eqnarray}
and fitted parameters are $A,B,C$. The fitting is carried out using MATHEMATICA package 
\begin{equation}
	A = 0.656585, B = 0.639077, C = -0.322505
\label{a3-03}
\end{equation}
after that we can define parameters $\alpha, r_0$ and $V^2_0$ 
\begin{equation}
	\alpha = 1.31954, \quad 
	r_0 \approx 2.01 \cdot 10^{-18} \; \text{cm}, \quad 
	V_0 \approx 32.25 \text{ Km} \cdot \text{s}^{-1}. 
\label{a3-13}
\end{equation}

\end{document}